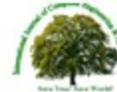

# OPTICAL DISK WITH BLU-RAY TECHNOLOGY

## T. Ravi Kumar[1], Dr. R. V. Krishnaiah[2]

*1(MTech-CS, D.R.K.Institute of science and technology, Hyderabad, India)*

*2(Principal, Dept of CSE, D.R.K Institute of Science and technology, Hyd, India.)*

## ABSTRACT

Blu-ray is the name of a next-generation optical disc format jointly developed by the Blu-ray Disc Association (BDA), a group of the world's leading consumer electronics, personal computer and media manufacturers. The format was developed to enable recording, rewriting and playback of high-definition video (HD), as well as storing large amounts of data. This extra capacity combined with the use of advanced video and audio codec's will offer consumers an unprecedented HD experience. While current optical disc technologies such as DVD, DVD±R, DVD±RW, and DVD-RAM rely on a red laser to read and write data, the new format uses a blue-violet laser instead, hence the name Blu-ray. Blu ray also promises some added security, making ways for copyright protections. Blu-ray discs can have a unique ID written on them to have copyright protection inside the recorded streams. Blu .ray disc takes the DVD technology one step further, just by using a laser with a nice color.

Keywords: Blu-ray Disc; BD; CD; DVD; HD-DVD; HDTV.

## I. INTRODUCTION

Blu-ray Disc format is required by the forthcoming of High Definition TV era, which calls for a new generation of optical storage after DVD. The storage demanding MPEG2 format HDTV normally has a data rate between 20 and 30 Mbps. For example, a 135 minutes recording of 25Mbps HDTV program requires a storage capacity about 24GB. Blu-ray Disc is specifically defined for such a capacity. To have the capacity five times that of DVD, BD employs the wavelength short as 405nm and numerical aperture high as 0.85. It is regarded the blue-violet wavelength will be the shortest wavelength in optical disc and 0.85 will also be the highest numerical aperture used for far-field optical storage. [6]To ensure a quick learning curve not only for the drive manufacturing but also for the disc production, Blu-ray Disc is designed to have even wider disc tolerances than those of DVD. Thanks for those more innovative concepts Blu-ray Disc is the most economical storage solution in terms of cost per giga byte. We can expect an even better result at a mass production level.

To introduce a new generation of optical storage, a long term standard is much more preferred rather than an interim solution. It is just like what happened in the history when the DVD format was defined. At that time, an interim solution of less capacity was proposed. Finally this solution was not introduced to the market.

The standards for 12-cm optical discs, CDs, DVDs, and Blu-ray rewritable discs (BD-RE Standard) were established in 1982, 1996, and 2002, respectively [1]. The recording capacity required by applications was the important issue when these standards were decided. The requirement for CDs was 74 minutes of recording 2- channel audio signals and a capacity of about 800 MB. For DVDs, the requirement as a video disc was the recording of a movie with a length of two hours and fifteen minutes using the SD (Standard Definition) with MPEG-2 compression. The capacity was determined to be 4.7 GB considering the balance with image quality.

The development of Fourier Optics in the middle and late 20th century provided an essential theoretical basis for the advent of optical storage technology [7]. The development during the same period in motion control technology, audio processing technology, error-correction coding technology, etc, enabled the standardization of disc storage to be first realized in the audio recording field, and commercialized at a low cost.

## II. LITERATURE REVIEWS

In 1983, compact disc technology was introduced into the United States revolutionizing the music and movie industries. Both music and movies benefited from the switch to digital which provided great clarity and storage space. "CD quality sound" became the catch phrase to denote quality music recording. As the industry grew, demand increased for better picture quality which meant greater storage capacity. DVDs were introduced to meet this need and were met with great success.

In the case of the Blu-ray *1) Disc, abbreviated as BD hereafter, a recording of an HDTV digital broadcast greater than two hours is needed since the BS digital broadcast started in 2000 and terrestrial digital broadcast has begun in 2003. It was a big motivation for us to realize the recorder using the optical disc. [1] In a DVD recorder, received and decoded video signals are compressed by an MPEG encoder and then recorded on the disc.

To record in the same fashion for an HDTV broadcast, an HDTV MPEG-2 encoder is required. However, such a device for home use has not yet been produced. In the case of BS digital broadcasts, signals are sent as a program stream at a fixed rate, which is 24 Mbps for one HDTV program. In the program stream of BS digital broadcast there is a case that the additional data stream is multiplexed, and it is desirable to record and read the data as is. Two hours of recording requires a recording capacity of 22 GB or more. This capacity is about 5 times that of DVDs, which cannot achieve this capacity by merely increasing their recording density.

To obtain this capacity we have developed a number of techniques such as: employing a blue-violet laser, increasing the numerical aperture of objective lens, making the optical beam passing substrate thin, 0.1 mm, and evenly thick, using an aberration compensation method of pickup adapted to the substrate thickness and dual

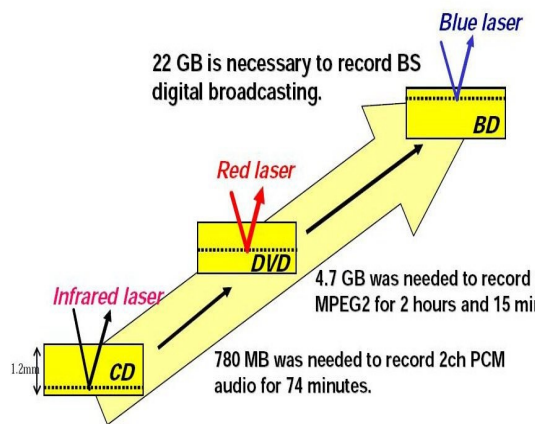

Fig1:  CD vs. DVD vs. BD

layer discs, improving the modulation method, enhancing the ability of the error correction circuit without sacrificing the efficiency, employing the Viterbi decoding method for reading signals and improving the S/N ratio and the inter symbol interference, using the on groove recording and highly reliable wobbling address system, developing high speed recording phase change media, etc. In addition, the convenient functions of a recording device have also been realized in the application formats.

These techniques are described in this paper. Furthermore, the key concepts of the Blu-ray standard such as the reason for employing 0.1 mm thick transparent layer and a dual layer recording disc will be described in each dedicated chapter. Following the rewritable system, the planning of a read-only system and write-once system has already started. In addition to high picture quality, the introduction of core and new functions is indispensable for the spread of the next generation package media. For example, during the switch from VHS to DVD, digital recording and interactive functions were newly introduced.[2] Consequently, it is anticipated that the specifications of BD-ROM will provide high performance interactive ness and a connection to broadband services, reflecting the demands of the movie industry.

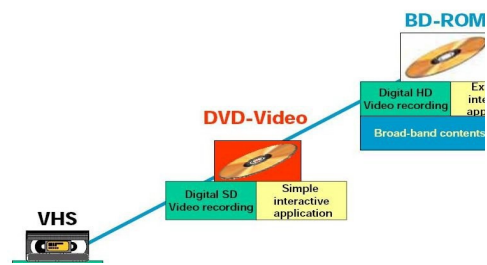

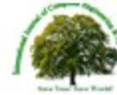

Fig2: Evolution of the package media for movie application

## III. BLU-RAY TECHNOLOGY

The technology utilizes a "blue" (actually blue-violet) laser diode operating at a wavelength of 405 nm to read and write data. Conventional DVDs and CDs use red and infrared lasers at 650 nm and 780 nm respectively. [1]

As a color comparison, the visible color of a powered fluorescent black light tube is dominated by mercury's bluish violet emissions at 435.8 nm.

The blue-violet laser diodes used in Blu-ray Disc drives operate at 405 nm, which is noticeably more violet (closer to the violet end of the spectrum) than the visible light from a black light. A side effect of the very short wavelength is that it causes many materials to fluoresce, and the raw beam does appear as whitish-blue if shone on a white fluorescent surface (such as a piece of paper). While future disc technologies may use fluorescent media, Blu-ray Disc systems operate in the same manner as D and DVD systems and do not make use of fluorescence effects to read out their data.

The blue-violet laser has a shorter wavelength than CD or DVD systems, and this shrinking makes it possible to store more information on a 12 cm (CD/DVD size) disc. The minimum "spot size" that a laser can be focused is limited by diffraction, and depends on the wavelength of the 11 light and the numerical aperture (NA) of the lens used to focus it. By decreasing the wavelength (moving toward the violet end of the spectrum), using a higher NA (higher quality) dual-lens system, and making the disk thinner (to avoid unwanted optical effects), the laser beam can be focused much tighter at the disk surface. This produces a smaller spot on the disc, and therefore allows more information to be physically contained in the same area. In addition to optical movements, Blu-ray Discs feature improvements in data encoding, closer track and pit spacing, allowing for even more data to be packed in.

**Principal of Operation**

When a diode is forward biased, holes from the p-region are injected into the n-region, and electrons from the n-region are injected into the p-region. If electrons and holes are present in the same region, they may radioactively recombine that is, the electron "falls into" he hole and emits a photon with the energy of the band gap. This is called spontaneous emission, and is the main source of light in a light-emitting diode.

Under suitable conditions, the electron and the hole may coexist in the same area for quite some time (on the order of microseconds) before they recombine. If a photon f exactly the right frequency happens along within this time period, recombination may be stimulated by the photon. This causes another photon of the same frequency to be emitted, with exactly the same direction, polarization and phase as the first photon.

In a laser diode, the semiconductor crystal is fashioned into a shape somewhat like a piece of paper very thin in one direction and rectangular in the other two. The crystal is n-doped, and the bottom is p-doped, resulting in a large, flat pn junction. The two ends of the crystal are cleaved so as to form perfectly smooth, parallel edges; two reflective parallel edges are called a Fabry-Perot cavity. Photons emitted in precisely the right direction will be reflected several times from each end face before they are emitted. Each time they pass through the cavity, the light is amplified by stimulated emission. Hence, if there is more amplification than loss, the diode begins to "laser".

**Hard-Coating Technology**

The entry of TDK to the BDF (as it was then), announced on 19 March 2004, was accompanied by a number of indications that could significantly improve the outlook for Blu-ray. TDK is to introduce hard-coating technologies that would enable bare disk (caddyless) handling, along with higher-speed recording heads and multi-layer recording technology (to increase storage densities).TDK's hard coating technique would give BDs scratch resistance and allow them to be cleaned of fingerprints with only a tissue, a procedure that would leave scratches on current CDs and DVDs.[3]

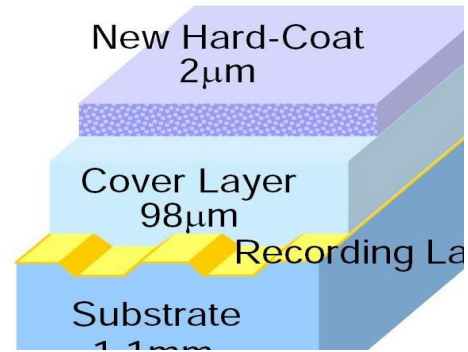

Fig3: Blu-ray disc cross section

**DISC STRUCTURE**

**Configuration of SL and DL Discs**

Figure shows the outline of a Single Layer BD Read-Only disc and Figure shows the outline of a Dual Layer BD Read-Only disc. To improve scratch resistance, the over layer can optionally be protected with an additional hard coat layer. One of the features that differentiate Blu-ray Disc from DVD recording systems is the position of the recording layer within the disc. For DVD, the recording layer is sandwiched between two 0.6-mm thick layers of plastic – typically polycarbonate.

The purpose of this is to shift surface scratches, fingerprints and dust particles to a position in the optical pathway where they have negligible effect - i.e. well away from the point of focus of the laser. However, burying the recording layer 0.6 mm below the surface of the disc also has disadvantages.

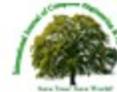

Due to the injection molding process used to produce them, disc substrates suffer from stress-induced birefringence, which means that they split the single incident laser light into two separate beams. If this splitting is excessive, the drive cannot read data reliably from the disc. Consequently, the injection molding process has always been a very critical part of CD and DVD production.[2] Another critical manufacturing tolerance, particularly for DVDs, is the flatness of the disc, because the laser beam becomes distorted if the disc surface is not perpendicular to the beam axis - a condition referred to as disc tilt. This distortion increases as the thickness of the cover layer increases and also increases for higher numerical To overcome these disadvantages, the recording layer in a Blu-ray Disc sits on the surface of a 1.1-mm thick plastic substrate, protected by a 0.1-mm thick cover layer.

With the substrate material no longer in the optical pathway, birefringence problems are eliminated. In addition, the closer proximity of the recording layer to the drive's objective lens reduces disc tilt sensitivity. This only leaves the problem of surface scratching and fingerprints, which can be prevented by applying a specifically.

**Single –Layer Disc**

Figure shows the outline of a Dual Layer BD Read-Only disc. To improve scratch resistance, the cover layer can optionally be protected with an additional hard coat layer. One of the features that differentiate Blu-ray Disc from DVD recording systems is the position of the recording layer within the disc.

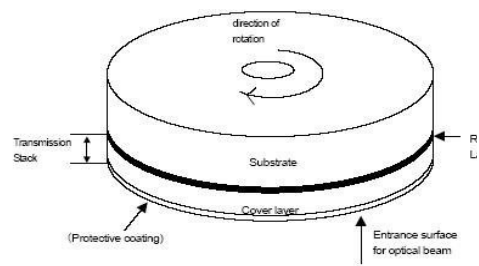

Fig4: Single-Layer Disc

**Dual Layer Disc**

Figure shows the outline of a Dual Layer BD Read-Only disc. To improve scratch resistance, the cover layer can optionally be protected with an additional hard coat layer. The different layers are shown. A spacing layer is used to separate the two information discs. Also the different transmission stacks are shown.

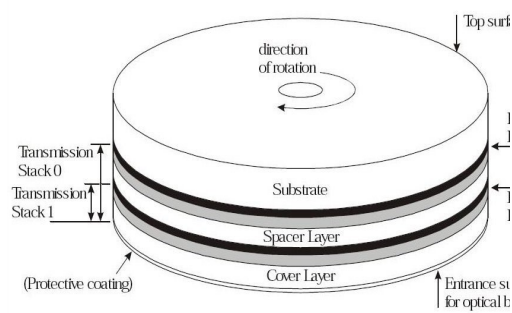

Fig5: Dual-Layer Disc

 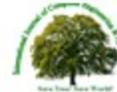

## IV. IMPLEMENTATION

The table below shows the technical specification of Blu-Ray

| Recording capacity: | 23.3GB/25GB/27GB |
|---|---|
| Laser wavelength: | 405nm (blue-violet las |

**Formats**

Blu-ray is initially designed in several different formats:

•**BD-ROM** (read-only) - for pre-recorded content

• **BD-R** (recordable) - for PC data storage

• **BD-RW** (rewritable) - for PC data storage

• **BD-RE** (rewritable) - for HDTV recording

| Lens numerical aperture (NA): | 0.85 |
|---|---|
| Data transfer rate: | 36Mbps |
| Disc diameter: | 120mm |
| Disc thickness: | 1 .2mm (optical transmittance protect layer: 0.1 mm) |
| Recording format: | Phase change recordi |
| Tracking format: | Groove recording |
| Tracking pitch: | 0.32um |
| Shortest pit length: | 0.160/0.149/0.138u |
| Recording phase density: | 16.8/1 8.0/1 9.5Gbit/in |
| Video recording format | MPEG2 video |
| Audio recording format: | AC3, MPEG1, Layer2, |
| Video and audio multiplexing format: | MPEG2 transport stre |

**Blu-Ray Vs VHS**

The Blu-ray Disc recorder represents a major leap forward in video recording technology as it enables recording of high-definition television (HDTV). It also offers a lot of new innovative features not possible with a traditional VCR:

• Random access, instantly jump to any spot on the disc

• Searching, quickly browse and preview recorded programs in real-time

• Create play lists, change the order of recorded programs and edit recorded Video

• Automatically find an empty space to avoid recording over programs

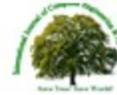

• Simultaneous recording and playback of video (enables Time slip/Chasing playback)

• Enhanced interactivity enables more advanced programs and games

• Broadband enabled, access web content, download subtitles and extras

• Improved picture, ability to record high-definition television (HDTV)

• Improved sound, ability to record surround sound (Dolby Digital, DTS, etc)

**Blu-Ray Vs Other Storage Devices**

The storage capacity of different digital storage technology varies a lot. A Usually used version of floppy disc has a capacity of 1.44MB while that of a CD is 700 MB & for DVD it is 4.7 GB. Also they have varying shell lives out of these DVD has the maximum. A DVD is very similar to a CD, but it has a much larger data capacity. A standard DVD holds about seven times more data than a CD does. This huge capacity means that a DVD has enough room to store a full length, MPEG-2-encoded movie, as well as a lot of other information. DVD can also be used to store almost eight hours of CD-quality music per side.[2] DVD is composed of several layers of plastic, totaling about 1.2 millimeters thick. Each layer is created by injection molding polycarbonate plastic.

**Comparison of BD AND DVD**

| Parameters | BD-ROM | DVD-ROM |
|---|---|---|
| Storage capacity (single-layer) | 25GB | 4.7GB |
| Storage capacity (dual-layer) | 50GB | 9.4GB |
| Laser wavelength | 405nm | 650nm |
| Numerical aperture (NA) | 0.85 | 0.60 |
| Protection layer | 0.1mm | 0.6mm |
| Data transfer rate | 36.0Mbps | 11.08Mbps |

A disc in the DVD format can currently hold 4.7 gigabytes of data. Unlike DVD technology, which uses red lasers to etch data onto the disc, the Blu-ray disc technology uses a blue-violet laser to record information.

The blue-violet laser has a shorter wavelength than the red lasers do, and with its Smaller area of focus, it can etch more data into the.[4] The digital information is etched on the discs in the form of microscopic pits. These pits are arranged in a continuous spiral track from the inside to the outside.

Using a red laser, with 650 nm wavelength, we can only store 4.7 GB on a single sided DVD. TV recording time is only one hour in best quality mode, and two, three or four hours with compromised pictures. Data capacity is inadequate for nonstop backup of a PC hard drive. The data transfer rate, around 10 Mbps, is not fast enough for high quality video.

## V. RESULT

Existing CD and DVD players and recorders will not be able to use Blu-Ray discs. New Blu-Ray players will need infra-red, red and blue lasers if they are also to play all kinds of CD and DVD recordings.

During the glass substrate making process, the substrate should be immersed in acetone first and remove the impurities and oil gas with ultrasonic vibrator. Then, take the substrate out of the acetone and cleanse in deionized water. After cleansing, the substrate should be placed in the oven to get rid of water vapor. The chromium of high reflectivity was chosen as reflecting layer because the substrate must possess the function of reflecting laser back to optical pickup head.

## VI. CONCLUSION

The advantages of BD are the capacity of reaching the far-field recording limit, large tolerance in processing, and the capability of the drive to be fully compatibility with CDs and DVDs in reading/writing, lowest storage cost per gigabyte. All these advantages make BD the first choice of storage in the high definition television era. The Blu-ray Disc Association has included almost all major companies in the optical storage industry, covering fields from content providing, disc processing, equipment manufacturing, product integration, to sale, and is forming a complete value chain.